\documentstyle[11pt,aaspp4,epsf]{article}

\def\0134{RX\,J0134-42} 
\def\G{$\Gamma_{\rm x}$ } 

\def\ros{{\sl ROSAT }} 
\def\asca{{\sl ASCA }}

\def\ein{{\sl Einstein }} 
\def\approxlt{\mathrel{\hbox{\rlap{\lower.55ex \hbox {$\sim$}}
        \kern-.3em \raise.4ex \hbox{$<$}}}}
\def\approxgt{\mathrel{\hbox{\rlap{\lower.55ex \hbox {$\sim$}}
        \kern-.3em \raise.4ex \hbox{$>$}}}}

\begin{document}
\title{The complex X-ray spectra of active galaxies with warm absorbers}
\author{Stefanie Komossa}
\affil{Max-Planck-Institut f\"ur extraterrestrische Physik, D-85740 Garching,
Germany}

\begin{abstract}
Warm absorbers are an important new probe of the
central regions of active galaxies (AGN).
So far, they revealed their existence mainly in the soft X-ray spectral region.
In observing and modeling this component, we can learn a lot about
the nature of the warm absorber itself, its relation to other components
of the active nucleus, and the intrinsic AGN X-ray spectral shape.

Here, we briefly review the basic X-ray spectral features of warm absorbers
(dust-free WAs, dusty WAs, peculiar $\sim$1.1 keV absorption, emission and reflection
components) and then discuss these in more detail based on analyses of individual
objects observed with the X-ray satellite {\sl ROSAT}. 
The importance of {\sl XMM} in improving our knowledge of the warm material
is discussed.

\end{abstract}

\section {Introduction} 

Warm absorbers reveal their presence by imprinting absorption
edges on the soft X-ray spectra of active galaxies (cf. Fig. 1).
These provide an important
diagnostic of the AGN central region.
The presence of an ionized absorber was first discovered in \ein observations
of the quasar MR 2251-178 (Halpern 1984).
With the availability of higher-quality soft X-ray spectra from \ros and \asca,
several more warm absorbers were found:
they are seen in about 50\% of the well studied Seyfert
galaxies (e.g., Nandra et al. 1993, Turner et al. 1993, 
%               N5548               N3783           
Weaver et al. 1994, Cappi et al. 1996, Mathur et al. 1997, Komossa et al. 1997c)
%      N4151          IC 4329A          N3516               N3786
as well as in some quasars (e.g. Fiore et al. 1993, Ulrich-Demoulin \& Molendi 1996,
Schartel et al. 1997).
More than one warm
absorber imprints its presence on the soft X-ray spectrum of MCG-6-30-15 
(e.g., Otani et al. 1996)
and NGC 3516 (Kriss et al. 1996). Particularly in these latter cases, improved
X-ray spectral resolution is required to clearly separate both components. 

Besides the absorption edges, the ionized absorber  
modifies the soft X-ray spectrum by emission lines,   
as emphasized by Netzer (1993). These lines contain important additional
information about the physical conditions of the ionized material.  
Whereas the detection of 
such lines was reported for NGC\,3783 (George et al. 1995), their
contribution (as calculated with the code {\em Cloudy}) was found to be
negligible for the best-fit warm absorbers in NGC\,4051 and NGC\,3227
(Komossa \& Fink 1997a,b). Emission features of the warm material
are best detectable against the reflected continuum component, i.e.
if the direct component is completely absorbed (cf. Fig. 1).  

In the last two years, evidence has accumulated that some warm absorbers
contain significant amounts of dust. This possibility was
suggested by Brandt et al. (1996)
to explain the lack of excess X-ray {\em cold} absorption despite strong optical
reddening 
of the quasar IRAS 13349+2438.
Explicit photoionization modelling of dusty warm gas using the code {\em Cloudy} 
showed that the modifications of the X-ray absorption spectrum can be quite
strong in the presence of dust, particularly for high column densities of the
warm absorber (Komossa \& Fink 1997a-d, Komossa \& Bade 1998).  

Another interesting recent development is the detection of spectral complexity
around 1.1 keV (e.g., Hayashida 1997).
`Standard' warm absorber models do not predict strong absorption features
around this energy; usually, the deepest edges are those of oxygen OVII and
OVIII at 0.74 and 0.87 keV, respectively. 
Among the suggested interpretations for the 1.1 keV feature is strongly
blueshifted oxygen absorption, implying relativistic outflow velocities 
(e.g., Leighly et al. 1997a). 

In the following, we first discuss two suggested candidates for dusty warm absorbers,
IRAS 13349+2438 and 4C+74.26, and predict the characteristic absorption features
observable with {\sl XMM}. 
We then investigate several scenarios to explain the
peculiar spectral features around 1.1 keV. 
The analyses are based on \ros X-ray observations and  photoionization calculations
carried out with the code {\em Cloudy} (Ferland 1993).

%---------------FIG---------------------------------------------------------
 \begin{figure}[h]
\epsfxsize=7.4cm
\epsfbox{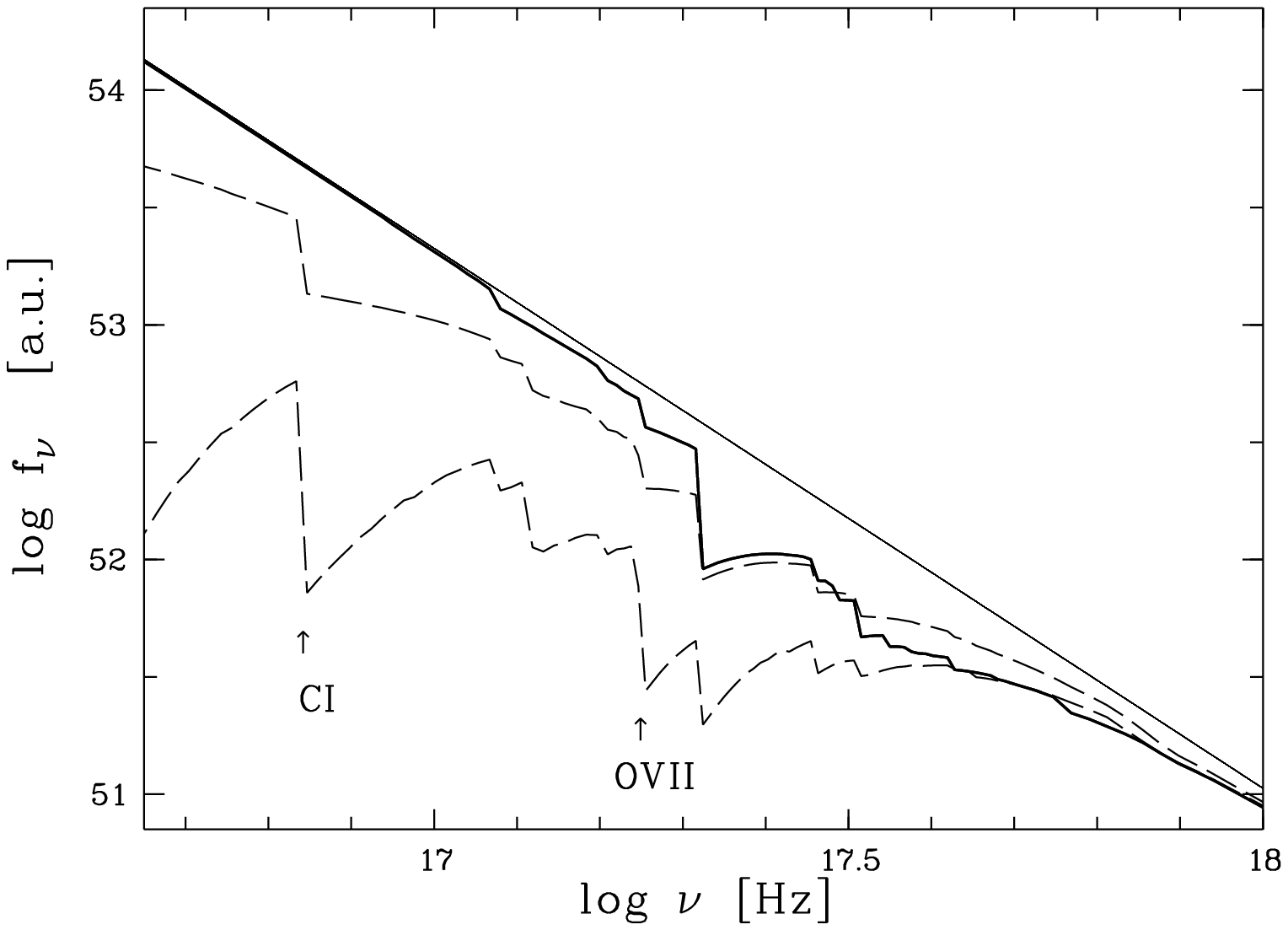}
    \vspace*{-5.25cm}\hspace*{7.9cm}
\epsfxsize=7.4cm
\epsfbox{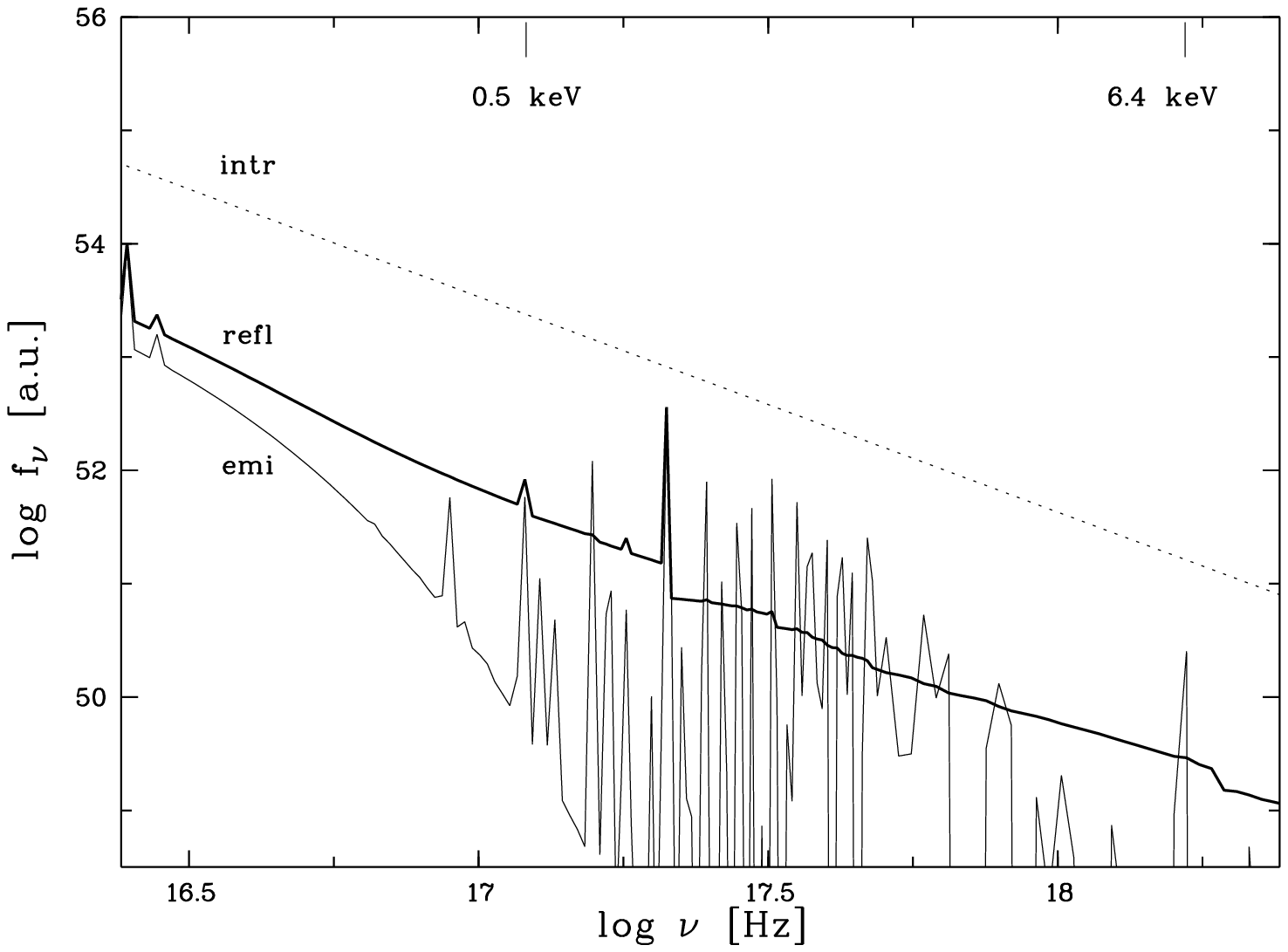}
 \caption[fig1]{{\bf Left:} {\small Changes of the X-ray absorption spectrum in the presence of dust.
The thin straight line represents the intrinsic continuum, the fat line shows
a dust-free warm absorber.
The dashed lines correspond to the same model after inclusion of dust
and depleted gas-phase metal abundances.
The dust was depleted relative to the standard Galactic-ISM
mixture by factors of 10 (upper dashed curve) and 3 (lower
dashed curve).
A characteristic feature of (the graphite species of)
dust is the strong edge of neutral carbon, labeled C\,I.   
{\bf Right:} Spectral components of warm material seen in emission and
reflection, calculated with Ferland's code {\em Cloudy} (see Komossa et al. 1998). 
The incident
continuum is shown as dotted line. The thin solid line corresponds to
the emitted spectrum and the thick solid line to the reflected spectrum.
The abscissa brackets the energy range 0.1 -- 10 keV.    } }
\label{fig1}
 \end{figure}
%--------------------------------------------------------------------------------

\section {Dusty warm absorbers} 

\subsection {IRAS\,13349+2438}
This quasar received a lot of attention, recently.
In X-rays, the presence of a dusty warm absorber was suggested (Brandt et al. 1996).
Here, we fit a model that explicitly includes the presence of dust to the \ros X-ray
spectrum. Although repeatedly suggested, such a model
has not been applied previously (for some results see Komossa \& Fink 1998).
Given the potentially strong modifications of the X-ray absorption spectrum
in the presence of dust,
it is important to  scrutinize whether a dusty warm absorber is consistent with
the observed X-ray spectrum.
Since some strong
features of dusty warm absorbers appear outside the \asca sensitivity range,
\ros data are best suited for this purpose; we used the pointed PSPC observation
of Dec. 1992.

In a first step, we fit a dust-{\em free} warm absorber (as in Brandt et al. 1996,
but using the additional
information on the hard X-ray powerlaw available from the ASCA observation,
$\Gamma_{\rm x}^{\rm 2-10 keV} \simeq -2.2$; Brinkmann et al. 1996, Brandt et al. 1997).
This gives an excellent fit with log $N_{\rm w}$=22.7 ($\chi^2_{\rm red}$ = 0.84).
If this same model is re-calculated by fixing $N_{\rm w}$ and
the other best-fit parameters but {\em adding dust} to the warm absorber
the X-ray spectral shape is drastically altered and the data can not be
fit at all ($\chi^2_{\rm red}$ = 150). This still holds if we allow for
non-standard dust, i.e., selectively exclude either the graphite or silicate
species.
It has to be kept in mind, though, that the expected column derived
from optical extinction is less than the X-ray value of $N_{\rm w}$ determined
under the above assumptions. Therefore, in a next step, we
allowed all parameters (except \G) to be free and checked, whether a dusty
warm absorber could be successfully fit at all.
This is not the case (e.g., if $N_{\rm w}$ is fixed to log $N_{\rm opt}$ = 21.2
we get $\chi^2_{\rm red}$ = 40).
The bad fit results can be partially traced back to the `flattening' effect of dust.
In fact, if we allow for a steeper intrinsic powerlaw spectrum, with \G $\simeq -2.9$
much steeper than the \asca value,
a dusty warm absorber with $N_{\rm w}=N_{\rm opt}$ fits the \ros
spectrum well ($\chi^2_{\rm red}$ = 1.2).
We also analyzed the \ros survey data and find the same
trends.  
At present,
there are several possible explanations for the {\sl ROSAT}-\asca spectral differences:
(i) variability in a {\em two}-component warm absorber, (ii) variability in
the intrinsic spectrum, or (iii) remaining {\sl ROSAT}-\asca
cross-calibration uncertainties.

\subsection {4C\,+74.26}
In an analysis of \ros and \asca data
of the radio-loud quasar 4C\,+74,
Brinkmann et al. (1998) find an unusually flat soft X-ray \ros spectrum
(\G $\simeq -1.3~{\rm to} -1.6$; as compared to \G $\simeq -2.2$ typically seen in     
radio-loud quasars), a steeper \asca powerlaw (PL) spectrum, and evidence for the
presence of a warm absorber.  
Applying the model of a {\em dusty} warm absorber to the \ros spectrum
we get a successful spectral fit, with a steeper intrinsic
PL spectrum (now consistent with the \asca value and the general expectation for
radio-loud quasars),  and a column density $N_{\rm w}$ 
consistent with optical reddening.
However, excess cold absorption provides an alternative
successful spectral fit, and higher X-ray spectral resolution 
is required to exclude a cold absorber.

\subsection {The role of XMM}

Besides IRAS\,13349, several more good candidates for dusty warm absorbers 
have been presented (NGC\,3227, Komossa \& Fink 1997b; NGC\,3786, 
Komossa \& Fink 1997c; MCG-6-30-15, Reynolds et al. 1997; IRAS\,17020+4544, 
Leighly et al. 1997b, Komossa \& Bade 1998)  which suggests this component to be
common in all types of AGN.
Signatures of the presence of dust are a carbon edge at 0.28 keV
and an oxygen edge at 0.56 keV, produced by inner-shell photoionization,
not yet individually resolved by current X-ray
instruments. Therefore, one main argument in favour of dust within the warm absorber
remained an indirect one: the discrepancy between the {\em large} amount of cold 
absorbing material inferred from optical data, and the {\em small or negligible}
excess cold absorption derived from the X-ray spectral analysis.   
The detection of the predicted absorption features with {\sl XMM} will therefore 
be an important and necessary confirmation of the existence of {\em dusty} warm
absorbers. It will then provide an interesting new approach 
to study the properties of dust in other galaxies, since the individual metal
absorption edges reflect the dust composition and amount of 
dust.{\footnote{Note that in case the dust is mixed with {\em cold} gas,
X-ray dust features would be extremely difficult to detect, since they are 
hard to distinguish from gas-phase absorption due to the same element;
the shift of the edge energy due to solid state effects 
is only of the order of a few eV (e.g., Greaves et al. 1984).}}

\section {Peculiar 1.1 KeV absorption: PG\,1404+226}

The \ros high-state and \asca spectrum of the luminous Sy galaxy PG 1404
show evidence for unexpectedly strong 1.1 keV absorption
(Ulrich \& Molendi 1996, Comastri et al. 1997, Leighly et al. 1997a).
Here, on the basis of detailed photoionization modelling of the absorbing
material under various conditions, we explore several 
scenarios to account for the 1.1 keV absorption
{\em without} invoking relativistic outflow: \\
 $\bullet$ a `standard' warm absorber of high ionization parameter, including a 
  strong contribution
 from emission and reflection; \\
 $\bullet$ non-solar abundances (over-abundant neon or under-abundant oxygen); \\
 $\bullet$ an additional soft excess overlapping with the absorption features
 (soft excess added to the {\em incident} continuum that illuminates the warm absorber). 

The following results are obtained: 
\noindent {\bf (i)} Models of high ionization parameter $U$ and/or with the emission and reflection
component added to the observed spectrum were calculated for a covering factor of 0.5.
In application to PG 1404, fitting such a model
improves
the quality of the fit, but the high-state data are still not
well matched. 
\noindent {\bf (ii)} One way to clearly change the depth of individual absorption edges, and
particularly to make the neon absorption dominate over oxygen in strength,
is a deviation from solar abundances, of either
overabundant neon or underabundant oxygen.
Several deviation factors were studied between an abundance of
up to O = 0.2 $\times$ solar and up to Ne = 4 $\times$ solar.
These models strongly improve the quality of the fit to PG\,1404
up to acceptable values (cf. Tab. 2 of Komossa \& Fink 1998). 
A potential problem for this model description, besides a difficulty
to explain strong deviations of O/Ne from the solar value in terms
of nucleosynthesis, is the location of the deepest edge.
In order to weaken sufficiently the oxygen absorption, a rather
high ionization parameter is required with the consequence that
the deepest neon edges are those of highly ionized species, around
1.36 keV, instead of 1.1 keV. 
\noindent {\bf (iii)}~Motivated by the \asca evidence for soft excesses in some NLSy1s,
a sequence of models was calculated with an additional hot
black body component of T = 0.1 keV. This component was included
in the ionizing SED that irradiates the absorber,
i.e. the change in
ionization structure of the warm material was self-consistently
calculated. 
This causes a complex spectral shape in the 1 keV region, with
the down-turning soft-excess and some Ne-K and Fe-L absorption
contributing to the X-ray features which then sensitively depend
on the strength of the soft excess (cf. Fig. 2 of Komossa \& Fink 1998).
(Note that the black body spectrum is {\em incident} on the warm material, and not
added afterwards as separate component; this causes a different ionization
structure and therefore soft X-ray spectral shape.)
A successful description of both, high- and low-state \ros data
is possible, but a rather broad edge structure is predicted.  \\
Data of high spectral resolution will be needed to discriminate between
these models, the interesting alternatives of relativistic outflow 
or high iron overabundance, or further scenarios to account for the 1.1 keV features.

\section {Summarizing conclusions}

Warm absorbers display many facets and may account for a variety
of observational phenomena, mostly   
in the X-ray spectral region. 
Their detailed study
will certainly
be a prime goal of the X-ray satellite {\sl XMM}.
In particular,
{\em dust}-created absorption edges, if confirmed, will play an important role not
only in probing components of the active nucleus, like the dusty torus,
but also are they a very useful new diagnostic   
of dust properties in other galaxies.

\end{document}